\documentclass[twoside,twocolumn,9pt]{article}
\usepackage{extsizes}

\usepackage[version=3]{mhchem}
\usepackage[left=1.5cm, right=1.5cm, top=1.785cm, bottom=2.0cm]{geometry}
\usepackage{balance}
\usepackage{sectsty}
\usepackage{graphicx} 
\usepackage{lastpage}
\usepackage[format=plain,justification=justified,singlelinecheck=false,font={stretch=1.125,small,sf},labelfont=bf,labelsep=space]{caption}
\usepackage{float}
\usepackage{fancyhdr}
\usepackage{fnpos}
\usepackage[english]{babel}
\addto{\captionsenglish}{%
  \renewcommand{\refname}{Notes and references}
}
\usepackage{array}
\usepackage{droidsans}
\usepackage{charter}
\usepackage[T1]{fontenc}
\usepackage[usenames,dvipsnames]{xcolor}
\usepackage{setspace}
\usepackage[compact]{titlesec}
\usepackage{hyperref}

\usepackage{epstopdf}

\usepackage{subcaption}
\usepackage{mathrsfs}
\usepackage{amsmath}
\usepackage{amssymb}
\usepackage{algorithm}
\usepackage{algorithmicx}
\usepackage[noend]{algpseudocode}

\usepackage{mathptmx}
\usepackage{mathrsfs}
\usepackage{bm}
\usepackage[usetitle=true, linkdoi=true]{rsc}

\DeclareMathAlphabet{\mathcal}{OMS}{cmsy}{m}{n}

\definecolor{cream}{RGB}{222,217,201}

\begin{document}

\pagestyle{fancy}
\thispagestyle{plain}
\fancypagestyle{plain}{
\renewcommand{\headrulewidth}{0pt}
}

\makeFNbottom
\makeatletter
\renewcommand\LARGE{\@setfontsize\LARGE{15pt}{17}}
\renewcommand\Large{\@setfontsize\Large{12pt}{14}}
\renewcommand\large{\@setfontsize\large{10pt}{12}}
\renewcommand\footnotesize{\@setfontsize\footnotesize{7pt}{10}}
\makeatother

\renewcommand{\thefootnote}{\fnsymbol{footnote}}
\renewcommand\footnoterule{\vspace*{1pt}%
\color{cream}\hrule width 3.5in height 0.4pt \color{black}\vspace*{5pt}} 
\setcounter{secnumdepth}{5}

\makeatletter 
\renewcommand\@biblabel[1]{#1}            
\renewcommand\@makefntext[1]%
{\noindent\makebox[0pt][r]{\@thefnmark\,}#1}
\makeatother 
\renewcommand{\figurename}{\small{Fig.}~}
\sectionfont{\sffamily\Large}
\subsectionfont{\normalsize}
\subsubsectionfont{\bf}
\setstretch{1.125} 
\setlength{\skip\footins}{0.8cm}
\setlength{\footnotesep}{0.25cm}
\setlength{\jot}{10pt}
\titlespacing*{\section}{0pt}{4pt}{4pt}
\titlespacing*{\subsection}{0pt}{15pt}{1pt}

\fancyfoot{}
\fancyfoot[RO]{\footnotesize{\sffamily{~\textbar  \hspace{2pt}\thepage}}}
\fancyfoot[LE]{\footnotesize{\sffamily{\thepage~\textbar}}}
\fancyhead{}
\renewcommand{\headrulewidth}{0pt} 
\renewcommand{\footrulewidth}{0pt}
\setlength{\arrayrulewidth}{1pt}
\setlength{\columnsep}{6.5mm}
\setlength\bibsep{1pt}

\makeatletter 
\newlength{\figrulesep} 
\setlength{\figrulesep}{0.5\textfloatsep} 

\newcommand{\topfigrule}{\vspace*{-1pt}%
\noindent{\color{cream}\rule[-\figrulesep]{\columnwidth}{1.5pt}} }

\newcommand{\botfigrule}{\vspace*{-2pt}%
\noindent{\color{cream}\rule[\figrulesep]{\columnwidth}{1.5pt}} }

\newcommand{\dblfigrule}{\vspace*{-1pt}%
\noindent{\color{cream}\rule[-\figrulesep]{\textwidth}{1.5pt}} }

\makeatother

\twocolumn[
  \begin{@twocolumnfalse}

\sffamily

\noindent\LARGE{\textbf{Illuminating the property space in crystal structure prediction using Quality-Diversity algorithms}} \vspace{0.6cm}

\noindent\large{Marta Wolinska\textit{$^{a}$}, Aron Walsh\textit{$^{b}$} and Antoine Cully\textit{$^{a}$}} \vspace{0.6cm}

\noindent\normalsize{The identification of materials with exceptional properties is an essential objective to enable technological progress. We propose the application of \textit{Quality-Diversity} algorithms to the field of crystal structure prediction. The objective of these algorithms is to identify a diverse set of high-performing solutions, which has been successful in a range of fields such as robotics, architecture and aeronautical engineering. As these methods rely on a high number of evaluations, we employ machine-learning surrogate models to compute the interatomic potential and material properties that are used to guide optimisation. Consequently, we also show the value of using neural networks to model crystal properties and enable the identification of novel composition--structure combinations. In this work, we specifically study the application of the MAP-Elites algorithm to predict polymorphs of TiO$_2$. We rediscover the known ground state, in addition to a set of other polymorphs with distinct properties. We validate our method for C, SiO$_2$ and SiC systems, where we show that the algorithm can uncover multiple local minima with distinct electronic and mechanical properties.
} 
\vspace{0.6cm}

 \end{@twocolumnfalse} \vspace{0.6cm}

]

\renewcommand*\rmdefault{bch}\normalfont\upshape
\rmfamily
\section*{}
\vspace{-1cm}

\footnotetext{\textit{$^{a}$~Adaptive and Intelligent Robotics Lab, Imperial College London, London SW7 2AZ, UK; a.cully@imperial.ac.uk}}
\footnotetext{\textit{$^{b}$~Thomas Young Centre and Department of Materials, Imperial College London, Exhibition Road, London SW7 2AZ, UK; a.walsh@imperial.ac.uk}}

\section{Introduction}

Inorganic crystals are an important class of materials, with their application spanning a range of applications, such as photovoltaic cells \cite{Green2014}, batteries\cite{MIZUSHIMA1980783} and transistors\cite{jian_high-performance_2020}. 
The computational discovery of new crystals is a promising avenue in identifying materials with the potential to augment our technological capabilities and accelerate progress in a range of fields. 

One of the main challenges of crystal structure prediction (CSP) is a search problem.
The search space of all possible crystals increases with $10^{N_{atoms}}$ \cite{oganov_crystal_2006}; this decreases to exponential if local relaxation is used \cite{oganov_crystal_2006, stillinger_exponential_1999}. 
As such, techniques that explore the space efficiently and effectively are required. 
This can be done using both data-driven or \textit{ab initio} techniques\cite{yin_search_2022}. 
Although data-driven techniques have been successful\cite{merchant_scaling_2023}, they rely on the availability of training data \cite{allahyari_coevolutionary_2020}, which becomes more limited as the complexity of systems increases.
Consequently, there are advantages to \textit{ab initio} techniques as they do not rely on a base knowledge of the chemical space. 
One such technique is evolutionary algorithms. 
Thanks to their inherent randomness they can explore a highly complex search space without getting stuck in local minima. 
They have been shown to work effectively in CSP \cite{glass_uspexevolutionary_2006, lonie_xtalopt_2011}. 

Novel technological applications often require materials with a combination of optimal (but likely conflicting) properties.
To access crystal structures exhibiting such unique and advanced properties, additional optimisation techniques are required. 
The associated computational search is constrained with limitations, such as availability of sufficient data for modelling and the size of the search space. 
This makes the ability to discover truly novel crystals and to select the right candidates for optimisation challenging\cite{allahyari_coevolutionary_2020}.
These limitations are further pronounced when optimising for multiple properties\cite{allahyari_coevolutionary_2020}.
One technique that considers multiple properties is multi-objective optimisation\cite{deb_multi-objective_2014, Allahyari2018}, which has been used effectively in a range of materials design studies\cite{omee_crystal_2023, Jablonka2021_bias_free}, including a combination of an evolutionary search with multiple objectives \cite{allahyari_coevolutionary_2020}. 
Multi-objective optimisation is typically designed to search for a set of solutions that lie at the trade-off of multiple conflicting objectives.
This however can be seen as a limitation, because potentially high-performing solutions that lie outside of this condition would be discarded \cite{pierrot_multi-objective_2022}. 
Such techniques are also not designed to explicitly provide diverse solutions to a problem, which could aid the user in understanding the feature space of their problem. 

To address these challenges we can use \textit{Quality-Diversity} (QD) algorithms -- an expanded framework built on top of evolutionary algorithms. 
They aim to provide a diverse set of high-performing solutions in some feature space, which effectively changes the optimisation objective. 
They allow the user to define a number of features of interest, which guide the optimisation to find diverse solutions while maximising overall fitness. 
In the context of CSP, the features could be any calculable material properties, while the fitness could be the energy function. 
Through optimisation they also provide the user with a better understanding of the feature space, which is why they are also referred to as \textit{illumination algorithms}.
These algorithms have been successfully used in a range of fields such as robotics to enable robots to learn new behaviours if they encounter damage \cite{cully_robots_2015}, in architecture to design buildings\cite{hagg_efficient_2023} or in aeronautics to design airfoils \cite{gaier_data-efficient_2017}. 

Given the success of QD in other fields, in this work we will apply one such algorithm to the problem of crystal structure prediction. 
A requirement of these techniques is a large number of evaluations. 
In CSP, a first-principles approach such as density functional theory (DFT) would typically be used to predict both the energy and properties of each structure.
As this approach is computationally expensive, it is not suitable for methods that require a high number of evaluations. 
However, in recent years machine learning surrogate models\cite{chen_learning_2021, chen_universal_2022, chen_graph_2019, deng_chgnet_2023}, which effectively model this energy function for a wide range of chemistries have been developed.
This is also true for material properties, as recorded in the MatBench benchmark\cite{riebesell_matbench_2023}.
The significant decrease in cost per evaluation for such surrogate models, creates an opportunity for new techniques to be tested and developed without the constraints of the number of evaluations required. 
This opportunity will be used in this work to demonstrate how QD algorithms can be used in crystal structure prediction. 
We demonstrate the capabilities of our proposed algorithm to generate a large collection of promising structures. 
We validated this on 4 materials to demonstrate that with minimal physical assumptions the algorithm uncovers multiple local minima with distinct electronic and mechanical properties.

\section{Background}
\subsection{Evolutionary Algorithms}
Evolutionary algorithms have been successfully used for crystal structure prediction, notably in well-established packages such as USPEX \cite{glass_uspexevolutionary_2006} and XtalOpt \cite{lonie_xtalopt_2011}.
They have gained popularity, not only because of their efficiency, but also because they can be based on first-principles quantum mechanical calculations. 
As such they are not constrained by data availability nor influenced by areas of the chemical space explored in past work\cite{allahyari_coevolutionary_2020}.

These algorithms follow the principles of evolution. 
First, some random solutions, known as \textit{individuals}, are generated and stored. 
An individual is defined by a set of \textit{genes}: in the context of crystal structures those would be the position of atoms in a cell as well as the cell size. 
Then a number of individuals are selected and their genes are randomly updated. 
Such updates are called \textit{mutations}; an example of a mutation in the context of crystal structures could be randomly adding Gaussian noise to the positions of atoms in a cell\cite{van_den_bossche_tight-binding_2018}. 

The quality of generated individuals is evaluated using a \textit{fitness function}. 
Based on this result individuals are either kept or discarded. 
This continues for a user-defined number of cycles, called \textit{generations}, or until a different exit condition is met.  

Evolutionary algorithms are a global optimisation technique, i.e. they aim to find the global minimum/maximum of a function. 
As such, they do not aim to provide an understanding search space, nor are they designed to search for multiple high-quality solutions.
QD algorithms are designed to address these challenges. 

\subsection{Quality-Diversity}
QD algorithms are an extension of evolutionary algorithms. 
They provide a framework that guides optimisation to find a diverse yet high-performing set of solutions. 
They have a two-fold objective: to globally maximise the diversity of solutions and to locally maximise their fitness. 
The following discussion presents a general view of QD algorithms, with the specific case of MAP-Elites introduced in the following section and summarised in Alg. \ref{alg:map_elites}.
As such, we will introduce the notation and reference Alg. \ref{alg:map_elites} in this section to facilitate interpretation. 

The key difference between evolutionary algorithms and QD is that an individual ($\theta$) is characterised not only using its fitness ($p$), but also using a \textit{feature vector} ($\bm{b}$). 
The feature vector contains any number of user-defined properties of interest.
An example feature vector of a crystal structure could contain its hardness and toughness. 
The feature vector is used to compare individuals to each other and to determine if newly generated individuals will be added to the set of stored solutions called an \textit{archive} ($\mathcal{A}$).
There are many benefits to adding the feature vector into optimisation these include the fact that (1) they illuminate the search space thus allowing its improved understanding without the constraints of multi-objective optimisation, (2) they are intuitive to use, (3) they are agnostic to the problem statement and (4) they can also be used at initial investigation to identify promising areas for exploration. 

The archive is stored within a \textit{container} ($\mathcal{C}$), which defines how the individuals are organised within the archive (lines 12-16 in Alg. \ref{alg:map_elites}).
A container can be as simple as a user-defined grid, where each cell is used to store an individual. 
When a new individual is generated, its feature vector is used to assign an individual to a cell. 
If the cell is empty, the individual is added to the archive, if it is not, the individual with the higher fitness is kept (line 15 Alg. \ref{alg:map_elites}) - thus ensuring fitness is locally maximised. 
The container does not have to be discretised by the user as is the case with a grid. 
A container can also be unstructured and dynamically determine the local environments, for instance using a minimum distance between \textit{k}-nearest neighbours.

\subsection{MAP-Elites}
An example of a state-of-the-art QD algorithm using a grid as the container is MAP-Elites (\textit{Multi-dimensional Archive of Phenotypic Elites}) \cite{mouret_illuminating_2015}. 
The goal of this algorithm is to obtain a set of solutions (archive), based on multiple dimensions defined by user-defined features (mutli-dimensional phenotypes), which contains the best (elite) solutions to a problem.

One of the strengths of this approach is that it can use any number of features.
However, as the number of dimensions increases if a constant resolution is maintained, the number of available cells within the grid increases exponentially. 
To avoid the curse of dimensionality, the following variation on the MAP-Elites algorithm can be used. 

CVT-MAP-Elites, uses Centroidal Vornoi Tesselation (CVT) to generate geometrically equivalent cells within any number of dimensions\cite{vassiliades_using_2017} (line 2 Alg. \ref{alg:map_elites}). 
This allows the user to define the desired number of cells in the container, while ensuring the same resolution across each feature.
Each cell within the resulting grid is referred to as a \textit{centroid}.
The CVT-MAP-Elites algorithm is reported in Alg. \ref{alg:map_elites}.
We employ CVT-MAP-Elites, but the shorthand MAP-Elites will be used for convenience as the two algorithms are equivalent aside from the container definition. 

\begin{algorithm}[ht]
\caption{CVT-MAP-Elites algorithm. Adapted from \cite{chatzilygeroudis_quality-diversity_2020, vassiliades_using_2017}}
\begin{algorithmic}[1]

\Procedure{MAP-Elites}{$[n_1, ..., n_d]$}
    \State $\mathcal{C}$ = CVT($[n_1, ..., n_d]$) \Comment{Run CVT and get centroids}
    \State $\mathcal{A} \leftarrow $create\_empty\_archive($[n_1, ..., n_d]$) 
    \For{i = 1 $\rightarrow G$ } \Comment{Initialisation: $G$ random $\theta$}
        \State $\theta $= random\_solution()
        \State \Call{add\_to\_archive}{$\theta$, $\mathcal{A}$}
    \EndFor
    
    \For{i = 1 $\rightarrow I$ } \Comment{Main loop, $I$ iterations}
        \State $\theta $= selection($\mathcal{A}$)
        \State $\theta'$ = variation($\theta$)
        \State \Call{add\_to\_archive}{$\theta$, $\mathcal{A}$}
    \EndFor
      \State \textbf{return} $\mathcal{A}$ 
\EndProcedure

\Procedure{add\_to\_archive}{$\theta$, $\mathcal{A}$}
    \State ($p, \bm{b}$) $\leftarrow$ evaluate($\bm{\theta}$)
    \State $c \leftarrow$ get\_index\_of\_closest\_centroid($\bm{b}$)
    \If {$\mathcal{A}(c) = null$ or $\mathcal{A}(c).p < p$}
        \State $\mathcal{A}(c) \leftarrow p, \bm{\theta}$
    \EndIf

\EndProcedure
\end{algorithmic}
\label{alg:map_elites}
\end{algorithm}

\begin{figure}[ht]
    \centering
    \begin{subfigure}{0.45\textwidth}
    \includegraphics[width=0.95\textwidth]{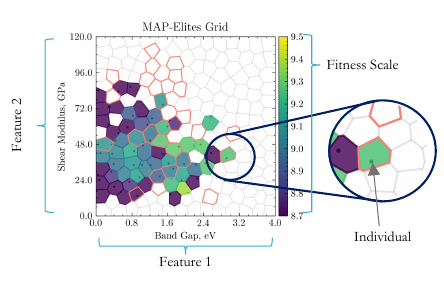}
    \caption{Annotated MAP-Elites Grid.}
    \label{fig:archive_interpretation_grid}
\end{subfigure}
\hfill
\begin{subfigure}{0.45\textwidth}
    \includegraphics[width=0.95\textwidth]{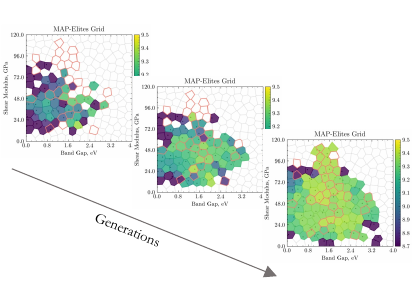}
    \caption{Sketch of optimisation over time.}
    \label{fig:archive_interpretation_generations}
\end{subfigure}
    \caption{Annotated MAP-Elites grid and representation of evolution in MAP-Elites solutions across generations.}
    \label{fig:archive_interpretation}
\end{figure}

To facilitate the interpretation of results a sample MAP-Elites grid is provided in Figure \ref{fig:archive_interpretation}. 
The x- and y-axis of the grid are described by two features, as annotated in Figure \ref{fig:archive_interpretation_grid}. 
The grid itself is defined using limits computed using CVT. 
A generated individual is represented using its feature vector in the grid using a scatter point. 
To aid visualisation the cell containing the individual is then coloured according to its fitness. 
For this work a known crystal system will be used, therefore cells where known structures would be positioned are marked with a red outline to aid interpretation.  

Figure \ref{fig:archive_interpretation_generations} visualises what an expected change could look like over some number of generations. 
We start with a smaller number of lower-performing (purple) solutions. 
Over the course of optimisation the number of solutions increases and they become increasingly high performing (yellow).

To quantify the performance of a run, \textit{coverage} and \textit{QD score} can be used alongside typical metrics, such as the maximum fitness. 
The coverage is the proportion of cells in the grid which contain solutions, thus capturing diversity.
The QD score is simply the sum of all individuals' fitness scores, which captures the global improvement in the quality of solutions. 

\section{Computational Methods}
\subsection{Computational Setup}
This work uses the CVT-MAP-Elites algorithm augmented with open-source Python packages for materials science.
Our implementation of CVT-MAP-Elites is based on the \texttt{pymap elites} implementation,\cite{mouret_illuminating_2015, vassiliades_using_2017} with some elements taken from the \texttt{QDAx} library\cite{chalumeau2023qdax}.
The starting structures were generated using the \texttt{pyxtal} generator \cite{fredericks_pyxtal_2021}. 
\textit{Strain} and \textit{Permutation} mutation operators from the \texttt{ase}\cite{hjorth_larsen_atomic_2017} library were used with a 50/50 chance of being selected\footnote{NB: For \ce{C} only the \textit{Strain} operator was used as only one type of atom was present in that experiment.}. 
The feature vector was formed of the shear modulus and band gap of a crystal. 
The absolute value of the energy per atom for each crystal was used as the basis of the fitness function\footnote{Since typically fitness is something to be maximised, we used the absolute value of energy rather than the negative values used by convention. We will therefore refer to maximising the absolute energy, which should be taken to be equivalent to minimising the energy in standard materials science nomenclature.}.
The energy was computed using \texttt{CHGNet}\cite{deng_chgnet_2023}, the band gap was computed using the \href{https://github.com/materialsvirtuallab/matgl/tree/main/pretrained_models/MEGNet-MP-2019.4.1-BandGap-mfi}{\texttt{MP-2019}} model from the Materials Graph Library, \texttt{matgl}\cite{chen_universal_2022, chen_learning_2021}, and the shear modulus was computed using the \texttt{\href{https://github.com/materialsvirtuallab/megnet/blob/master/mvl_models/mp-2018.6.1/README.md}{MP-2018}} model from \texttt{MEGNET}\cite{chen_graph_2019, chen_learning_2021}. 

There is a range of hyperparameters that are required both from QD and CSP perspectives. 
The MAP-Elites grid was discretised into 200 cells, random structures were generated in batches of 20 until a minimum of 10\% of the grid was populated with individuals prior to starting mutations. 
When an individual was generated it was relaxed for up to 100 steps with a maximum force tolerance for relaxation of 0.2 eV atom$^{-1}$. 
Then, at each generation 100 individuals were selected for mutation; this means that some individuals were mutated multiple times at each generation.
We primarily used \ce{TiO2} with 24 atoms per cell. 
This system was selected due to its polymorphic nature and it has been used as a benchmark material\cite{falls_xtalopt_2021, lonie_xtalopt_2011, lyakhov_new_2013}. 

The cell was initialised with a volume of $450$ \AA, and the scaling factor between inter-atomic distances was set to 0.4. 
The unit cell border lengths were set to $2-60$\AA, and the maximum angles were set to $0-\pi$. 
Maximum angles between unit cell vector and the plane created by the other two vectors are $[20^\circ, 160^\circ]$. 
The CSP-driven hyperparameters were set based on previous work in the field and where relevant they were set to be less 
restrictive\cite{falls_xtalopt_2021, lyakhov_new_2013, hjorth_larsen_atomic_2017, lonie_xtalopt_2011, van_den_bossche_tight-binding_2018}. 
The comparison against related work is provided as electronic supplementary information (ESI). 

\subsection{Fitness Function}
We introduce an additional consideration into the fitness function. 
To ensure that realistic and stable solutions were added to the archive, we first evaluated the maximum force acting within a structure. 
This is equivalent to the computation done within \texttt{CHGNet} during each step of relaxation. 

As summarised in Equation \ref{eq:force_threshold}, if the maximum force acting on a structure was higher than the preset threshold $F$, the fitness was set to the negative of the maximum force. 
If the maximum force was lower than the threshold, then the absolute value of the energy was used. 
Thus first we minimised the force to ensure realistic solutions would compete with each other to find the best absolute energy. 

A key benefit of this technique is that it provided a way to filter out unrealistic solutions that exploited out-of-distribution behaviour of \texttt{CHGNet} resulting in unrealistic energies exceeding hundreds or thousands of eV atom$^{-1}$. 
As such thanks to the force threshold, experiments can be run more reliably. 
The threshold was set to 9 eV \AA$^{-1}$. 
This was determined by computing the maximum force on the reference \ce{TiO2} structures and setting a value that would capture all of them (figure available in ESI). 

\begin{equation}
    p =
    \begin{cases}
        -1 * |max(\frac{\partial E}{\partial \theta}) - F| & \text{if } max(\frac{\partial E}{\partial \theta}) > F\\
        E(\theta) & \text{if } max(\frac{\partial E}{\partial \theta}) <= F
    \end{cases}
    \label{eq:force_threshold}
\end{equation}
\begin{center}
    \footnotesize 
where $E(\theta)$ is the energy function,  $max(\frac{dE}{dx})$ is the maximum force acting on a structure, $F$ is the value  of the preset threshold.
\end{center}

\subsection{Algorithm Evaluation}
Once the algorithm run is completed, all structures in the archive were compared against known polymorphs sourced from the Materials Project\cite{jain_commentary_2013}. 
The structures were evaluated using: 
(1) the \texttt{StructureMatcher} class from \texttt{pymatgen}\cite{ong_python_2013}; 
(2) by comparing the space group symmetry of the generated structures;
(3) by computing the fingerprint distance using the \texttt{ase}\cite{hjorth_larsen_atomic_2017} implementation. 

Based on a combination of these metrics a confidence level was assigned if a match was found. 
\texttt{StructureMatcher} was used as the primary metric. 
A \textit{gold standard} confidence was assigned if a match was found and it was in the right centroid within the feature space. 
If there was a match but the centroid did not match we assigned a \textit{high} confidence. 
A \textit{medium} confidence was assigned if the other two metrics were met or if either was met and the structure was in the same centroid as the reference.
A \textit{low} confidence was assigned if only one of the other two metrics was met. 
Otherwise, \textit{no match} was assigned to a generated structure.

Here it is important to note the limitations stemming from the evaluation method and the underlying property prediction models.
Firstly, let us consider the fact that the surrogate models are trained on realistic structures. 
As such, similar yet slightly perturbed structures could be evaluated to have different properties, thus assigning them to different centroids.
Secondly, the comparison of generated structures with reference structures is limited by the tolerances used by the evaluation methods. 
As these were set to be quite tolerant, dissimilar structures could be evaluated to be equivalent.
This means that multiple, ultimately equivalent, structures can be generated and assigned to different centroids or that the structures that are considered equivalent in this work are in fact distinct. Consequently, in this work MAP-Elites will be limited in the number of structures it can find.
The final diversity of the archive could thus be improved implementing techniques that address these constraints, such as periodically removing random individuals from the archive or fine-tuning the evaluation models.

\section{Results and Discussion}
Three experiments are reported to attain the following objectives.
Firstly, known reference structures of \ce{TiO2} were plotted within a MAP-Elites grid to validate that a wide range of structures with differing properties could be discovered. 
This also illustrates what a result would look like if all references available to be discovered within this resolution were found. 
Secondly, we demonstrate that using MAP-Elites multiple known structures of \ce{TiO2} are consistently found in a single algorithm run. 
Lastly, our method is applied to three other systems to demonstrate its versatility.

\subsection{Reference Data -- Titanium Dioxide}
\begin{figure}
    \centering
    \includegraphics[width=0.35\textwidth]{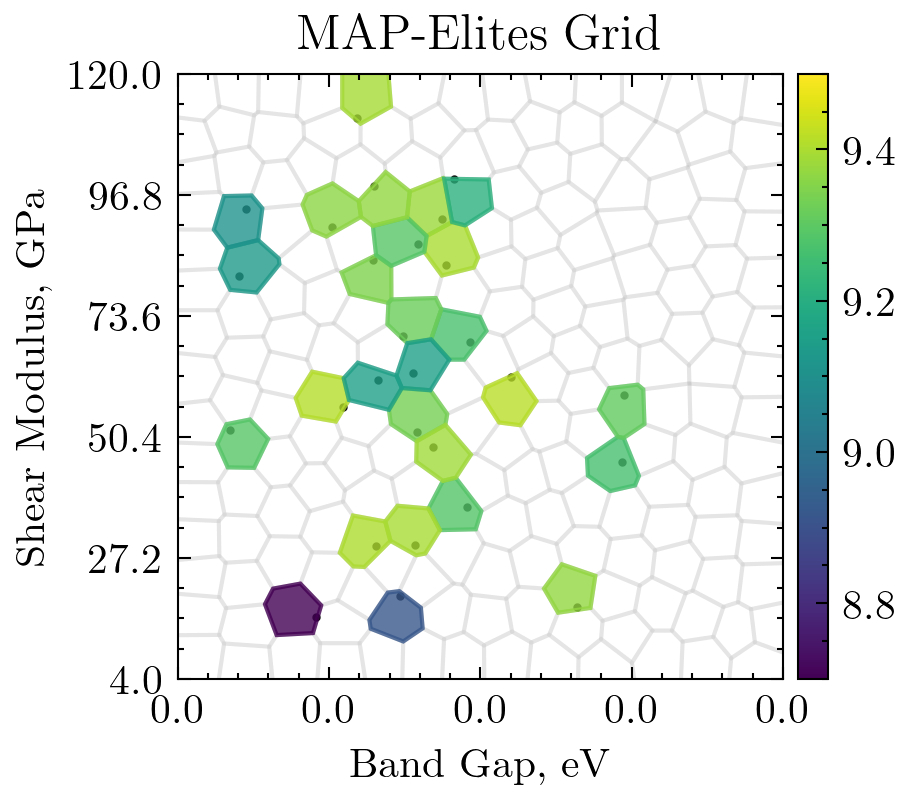}
    \caption{Known reference structures of \ce{TiO2} with 24 atoms or fewer plotted in a MAP-Elites grid.}
    \label{fig:reference_tio2}
\end{figure}

The known crystal structures, as sourced from the Materials Project, were plotted in a MAP-Elites grid in Figure \ref{fig:reference_tio2}.
To allow a wide breadth of solutions both theoretical and experimentally observed structures are included.

There are 34 polymorphs which contain 24 atoms or fewer, of these 8 are reported to be experimentally observed. 
In Figure \ref{fig:reference_tio2} however, only 29 centroids are filled.
Due to the resolution of the grid, some reference solutions were forced to compete and only the fitter structures (higher absolute energy) were kept. 
We can observe that the known reference structures are distributed across a wide range of band gap and shear modulus values, with no apparent correlation between them.
The availability of a wide range of structures not correlated within the feature space validates the suitability of the system and properties for this case study.

\subsection{Results -- Titanium Dioxide}
\begin{figure}
    \begin{subfigure}{0.45\textwidth}
        \centering
        \includegraphics[width=0.8\textwidth]{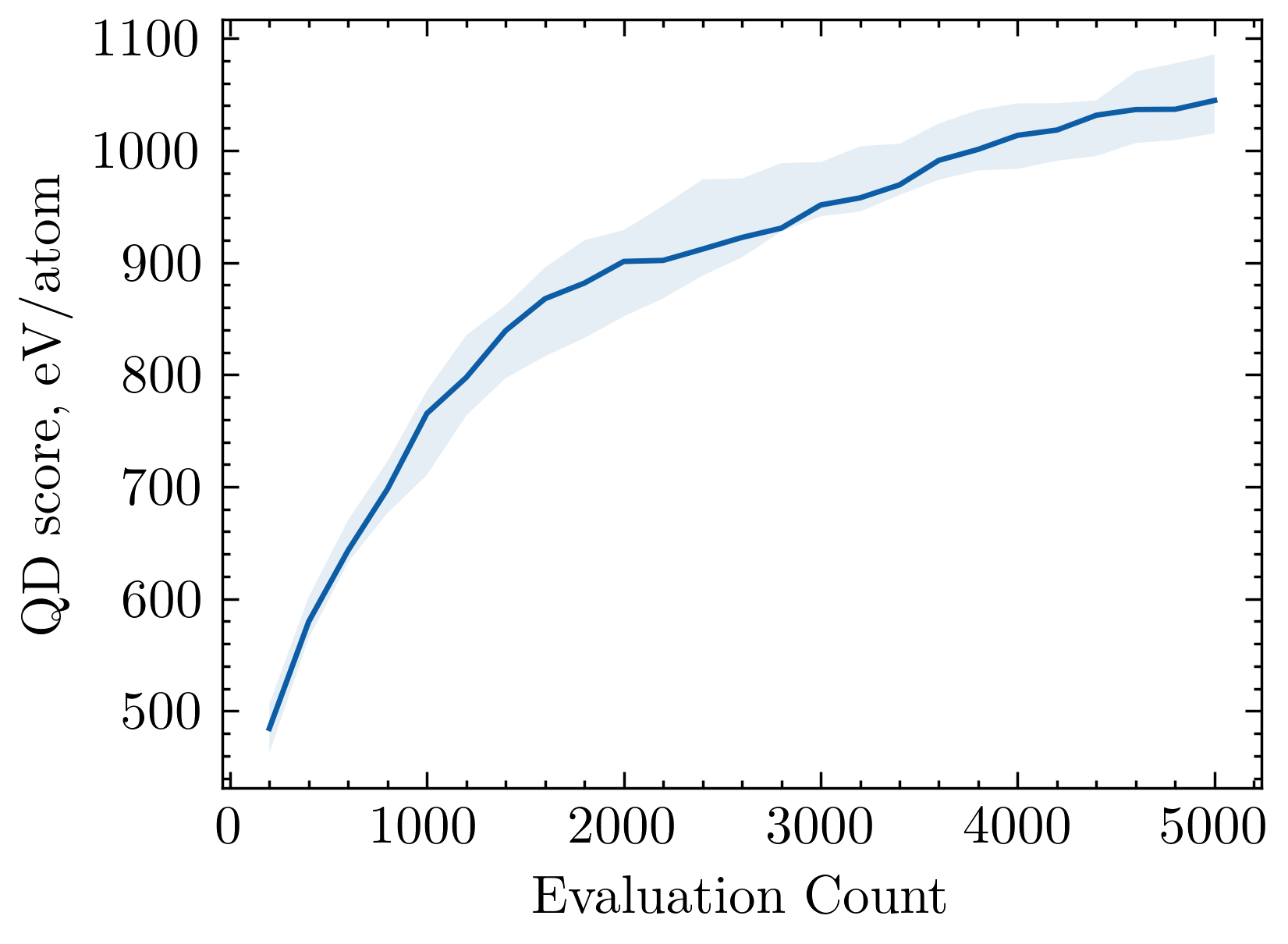}
        \caption{QD score over 5000 generations.}
        \label{fig:tio2_qd_score}
    \end{subfigure}
\hfill   
    \begin{subfigure}{0.45\textwidth}
        \centering
        \includegraphics[width=0.8\textwidth]{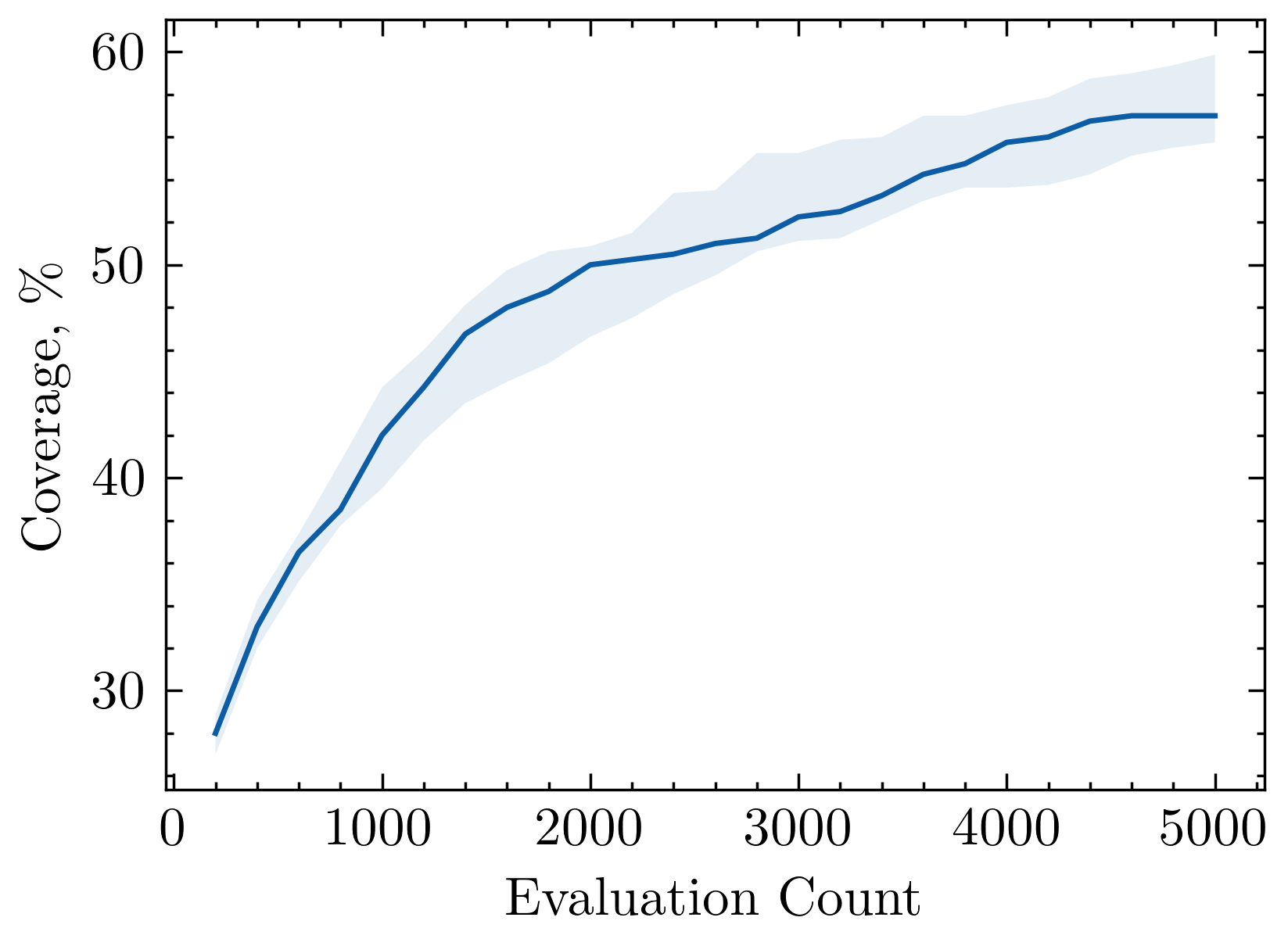}
        \caption{Coverage over 5000 generations.}
        \label{fig:tio2_coverage}
    \end{subfigure}  
\hfill
    \begin{subfigure}{0.45\textwidth}
        \centering
        \includegraphics[width=0.8\textwidth]{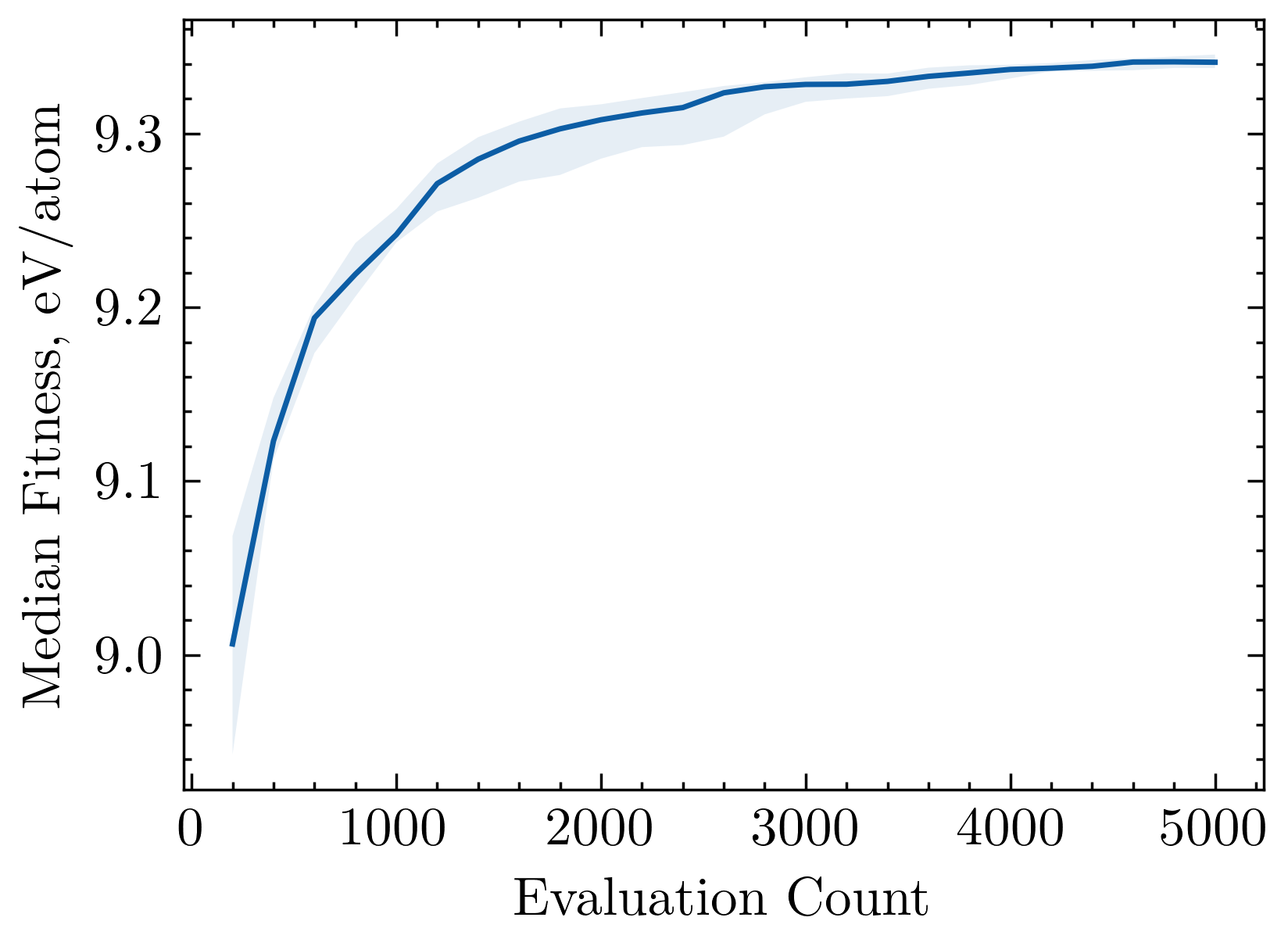}
        \caption{Median fitness over 5000 generations.}
        \label{fig:tio2_median_fitness}
    \end{subfigure}  
\hfill
    \centering
    \begin{subfigure}{0.45\textwidth}
        \centering
        \includegraphics[width=0.8 \textwidth]{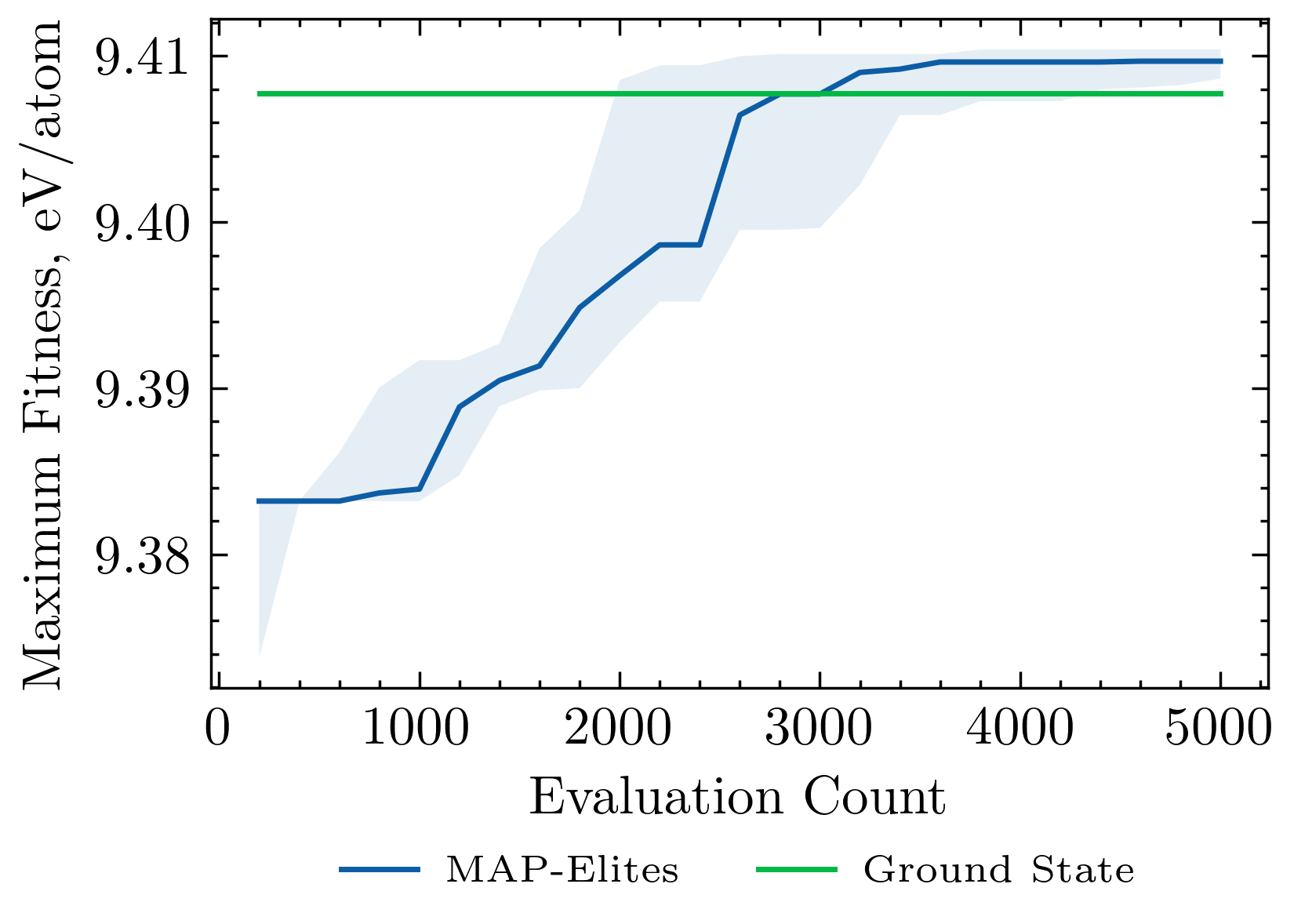}
        \caption{Maximum fitness over 5000 generations.}
        \label{fig:tio2_max_fitness}
    \end{subfigure}
    
    \caption{Median values of QD score, coverage, median fitness and maximum fitness averaged on 10 experiments across 5000 evaluations. The shaded are represents the $25^{th}$ and $75^{th}$ percentiles.}
    \label{fig:tio2_statistics}
\end{figure}

The results from the application of the MAP-Elites algorithm were averaged across 10 experiments. 
We observed that in all 10 experiments the global minimum was found, and that on average 10 (with a standard deviation of 2.2) other unique matches were found. 
On average 3 (with a standard deviation of 1.2) of the unique matches were found with a high or gold standard confidence. 
This demonstrates that the global minimum can be found in a reproducible manner alongside multiple other structures. 
To understand some of the dynamics of applying MAP-Elites to this crystal structure prediction problem, comparative statistics are plotted in Figure \ref{fig:tio2_statistics}.

The increase in QD score, as shown in Figure \ref{fig:tio2_qd_score}, shows that overall the fitness of the archive is increasing, which means that there are more high-performing solutions present overtime. 
This can be both because centroids are populated with increasingly high-performing solutions, and that more solutions are being added to the archive. 
The increase in the overall number of solutions is demonstrated by the increase in coverage as shown in Figure \ref{fig:tio2_coverage}. 
The increase in the median fitness of individuals is shown in Figure \ref{fig:tio2_median_fitness}.

In Figure \ref{fig:tio2_max_fitness}, we observe that the maximum attained fitness increases with the number of evaluations, indeed attaining the value of the ground state. 
The algorithm effectively explores the energy function to find its global minimum. 
This is expected since evolutionary algorithms have been extensively used for this application, but provides confidence in the algorithm setup, surrogate models used and hyperparameters set in this work. 

\begin{figure}
    \centering
    \includegraphics[width=0.35\textwidth]{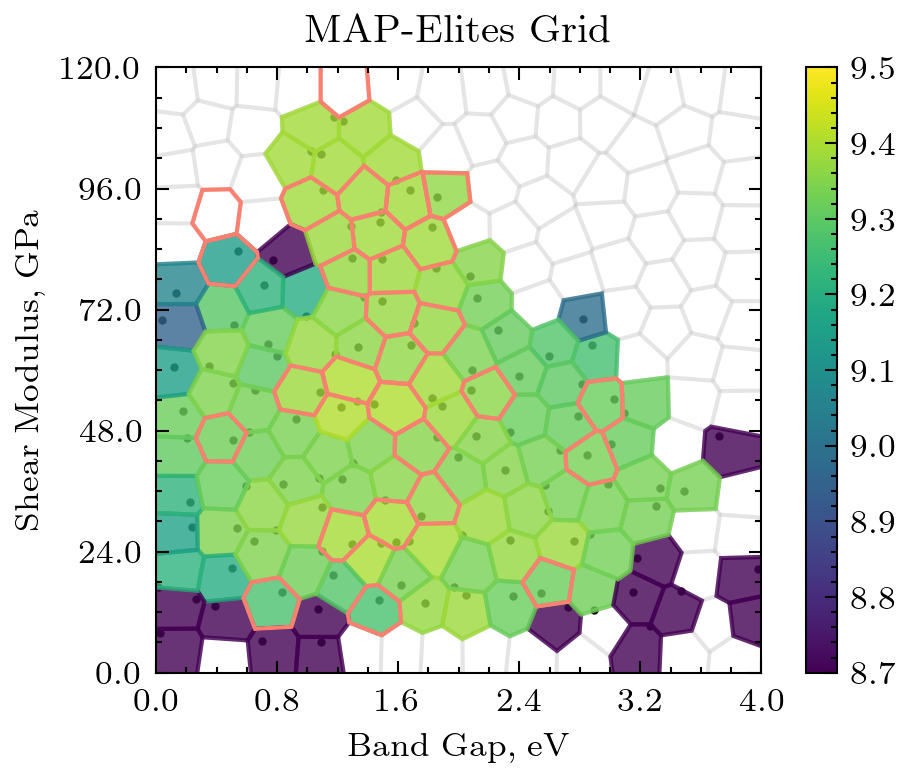}
    \caption{Sample archive after 5000 evaluations. Centroids where reference solutions are expected are marked with a red outline.}
    \label{fig:tio2_sample_archive}
\end{figure}

These metrics do not capture whether the expected reference structures are generated, nor how different the generated structures are from each other. 
Therefore a sample archive is inspected below. The archive was randomly selected from 10 experiments excluding the 2 best performing and 2 worst performing archives.
Figure \ref{fig:tio2_sample_archive} shows the sample MAP-Elites grid after 5000 evaluations. 
We can observe that the majority of the archive is populated with high-performing (yellow) individuals.
Additionally, all but 2 centroids where reference structures are expected are populated. 
The algorithm is effectively exploring the areas of the feature space where solutions are expected. 

\subsubsection{Identification of Reference Structures}
\begin{figure}
    \centering
    \begin{subfigure}{0.45\textwidth}
        \centering
        \includegraphics[width=0.7\textwidth]{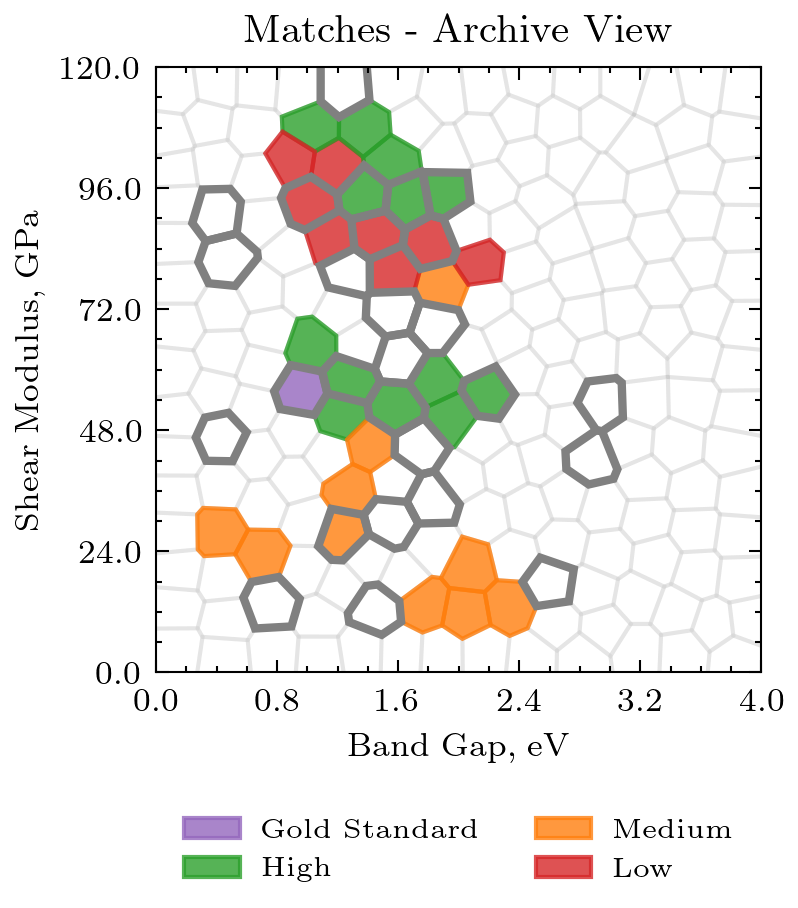}
        \caption{Archive with centroids containing a match to a reference structure highlighted.}
        \label{fig:tio2_archive_matches}
    \end{subfigure}
\hfill
    \begin{subfigure}{0.45\textwidth}
        \centering
        \includegraphics[width=0.7\textwidth]{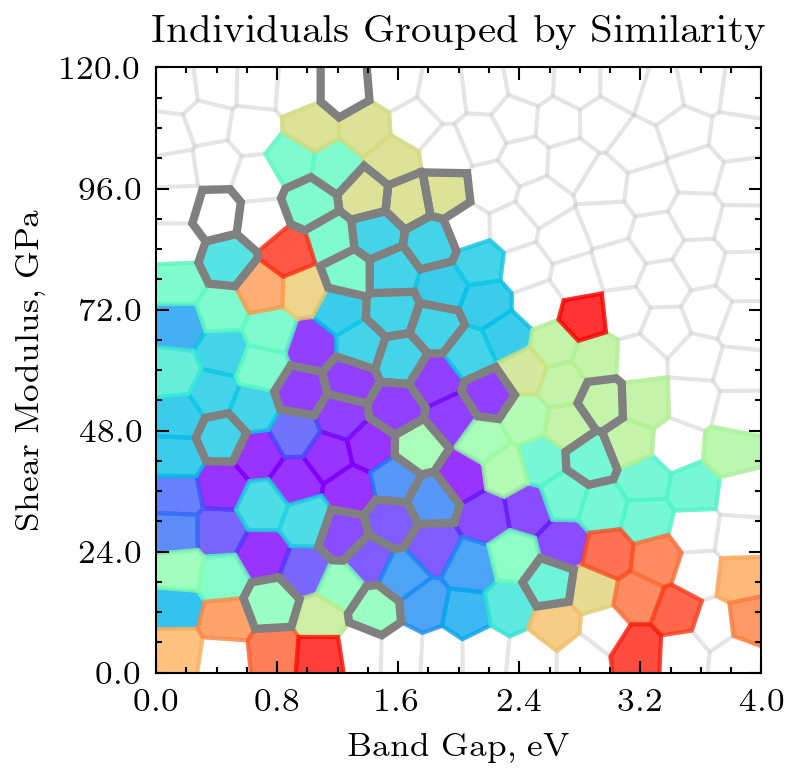}
        \caption{Archive view of structures grouped by similarity using \texttt{pymatgen} \texttt{StructureMatcher} class \cite{ong_python_2013}. Similar solutions are represented using the same colour. To aid visualisations the groups are numbered in an arbitrary order.}
        \label{fig:tio2_structure_similarity}
    \end{subfigure}
\hfill   
    \begin{subfigure}{0.45\textwidth}
        \centering
        \includegraphics[width=0.7\textwidth]{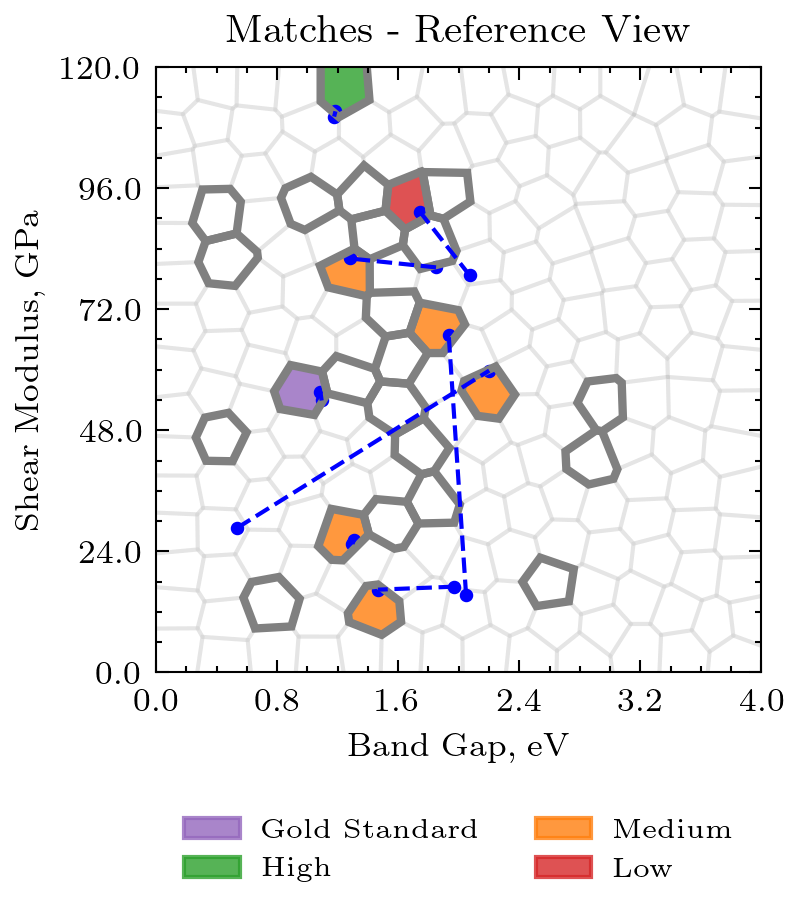}
        \caption{Unique structure matches found during optimisation.}
        \label{fig:tio2_unique_matches}
    \end{subfigure}
    \caption{Visual analysis of behaviours within sample archive.}
    \label{fig:tio2_archive_views}
\end{figure}

To understand whether centroids are populated with expected polymorphs, the generated structures were analysed.
Where a match was identified, the centroid was coloured using the confidence level of the match, as reported in Figure \ref{fig:tio2_archive_matches}. 
It is possible that by exploiting out-of-distribution behaviours multiple centroids are populated by similar structures (e.g. slightly distorted versions of a single polymorph).
This is confirmed by Figure \ref{fig:tio2_structure_similarity}, where similar structures were grouped using the same colour. 
We can see that there are indeed groups of solutions that take up multiple centroids in the map but that overall there is a wide range of dissimilar solutions generated. 
In this experiment, 49 groups were identified. 

To remove the duplication of structures, Euclidean distance in the feature space between the position of the reference and generated solutions was computed, and the structure with the shortest distance was kept. 
The results are visualised in Figure \ref{fig:tio2_unique_matches}. 
The centroid where the reference structure should be found is coloured using the confidence level. 
The positions of the reference and generated structures are indicated by the scatter points connected via the dashed line.
For the majority of the structures, the position of the two structures in the feature space is similar, but it is possible for them to lie quite far apart.
This can be caused by the fact that the underlying models are sensitive to the structure's definition. 

\begin{figure*}[h]
    \centering
    \includegraphics[width=\textwidth]{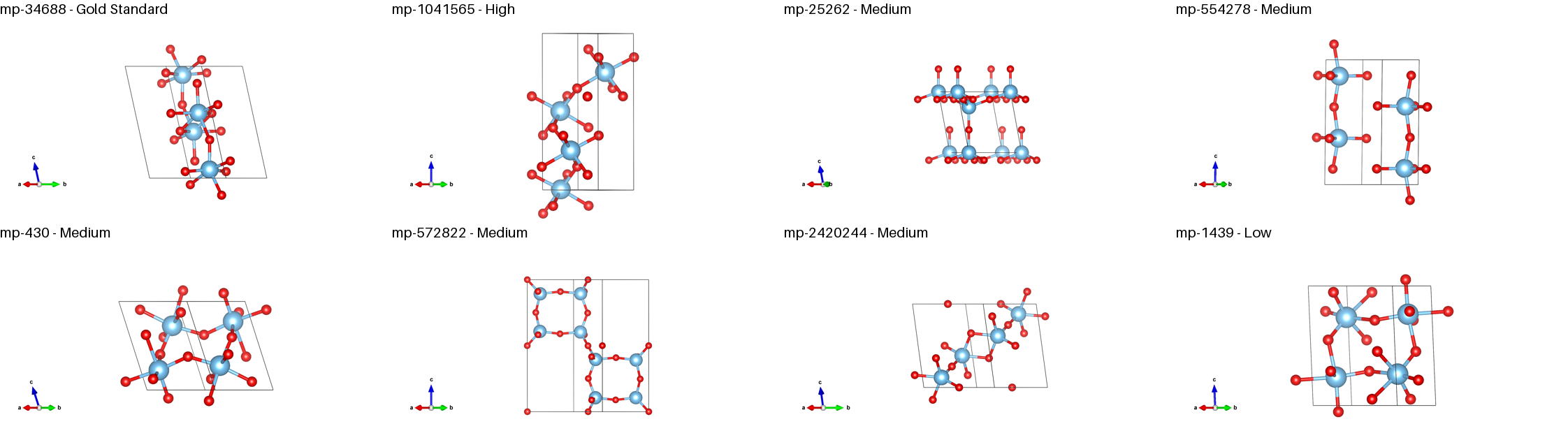}
    \caption{Crystallographic unit cells of the \ce{TiO2} polymorphs generated during optimisation. Structures were reduced to their primitive cell using \texttt{pymatgen} \texttt{SpacegroupAnalyzer} utility \cite{ong_python_2013}, visualised using VESTA\cite{vesta}. Each system is labelled by the corresponding Materials Project ID and confidence of the match.    
    }
    \label{fig:tio2_crystals}
\end{figure*}

The structures identified are visualised in Figure \ref{fig:tio2_crystals}.
We can observe that as a whole the structures are chemically realistic and diverse in terms of connectivity in the atomic building blocks. 
However, some structures share common motifs.
This is expected because, as highlighted above, within the tolerances used in this work, some reference structures are considered equivalent. 
This is the case for instance for the ground state (\textit{mp-390}) and a theoretical structure \textit{mp-34688}. 
The supporting analysis can be found in the ESI.

\subsubsection{Exploration Considerations}
\begin{figure}
    \centering
    \includegraphics[width=0.4\textwidth]{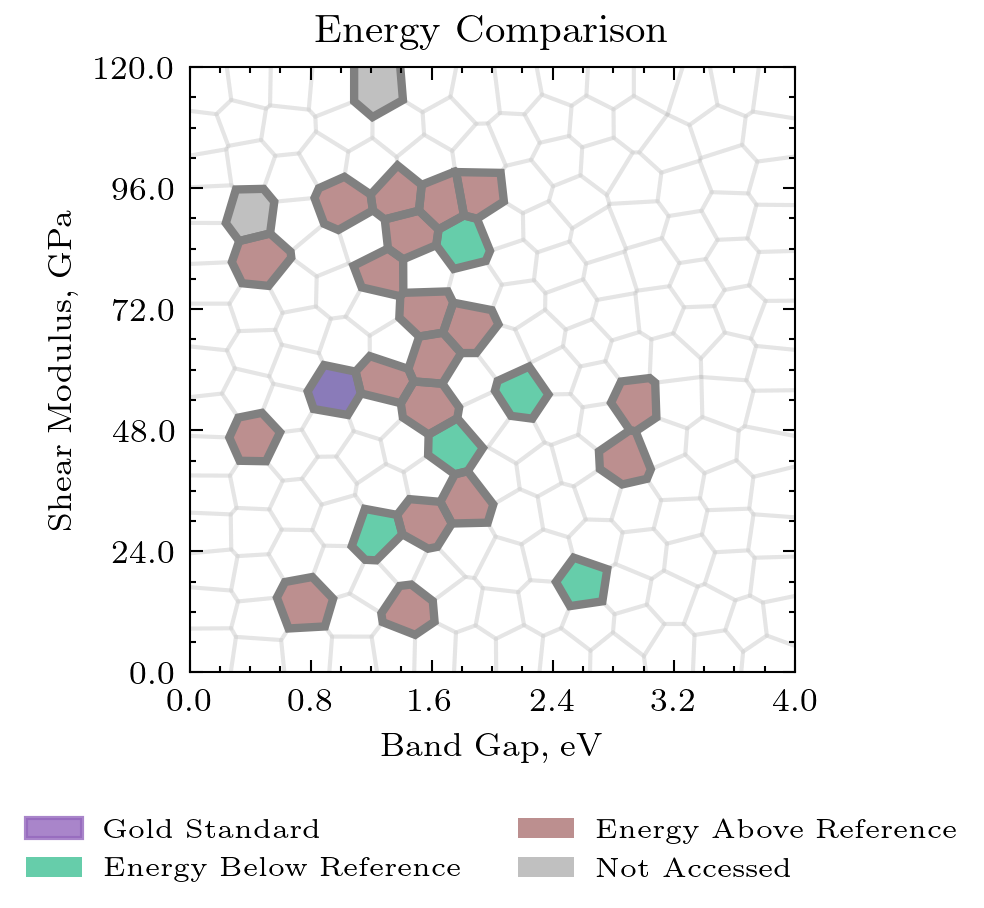}
    \caption{Energy difference between generated structure allocated to centroid and expected reference structure.}
    \label{fig:tio2_energy_difference}
\end{figure}

To confirm whether the algorithm is indeed attaining the expected values of absolute energy in each centroid, the difference in absolute energy between the generated and reference solutions in each centroid was computed. 
This enables us to understand the exploration of the algorithm. 
The results are plotted in Figure \ref{fig:tio2_energy_difference}. 
If the energy is higher than that of the reference this means that even if the reference structure had been discovered during optimisation, it would have been rejected due to its worse energy. 
This is the case for the majority of the reference centroids as shown in Figure \ref{fig:tio2_energy_difference}, which allows us to make three hypothesis: (1) more stable solutions could have been identified during optimisation than those present in the final archive, (2) the same base structure with small changes can be used to fool the neural networks into predicting varying property values and (3) the algorithm exploits unexplored areas of the neural network to fool it into predicting high absolute energy values. 
The last observation could be a powerful tool in generating structures to train more robust surrogate models for crystal structure prediction. 
Indeed QD methods have been used in past work exactly for this purpose \cite{grillotti_unsupervised_2022, grillotti_relevance-guided_2022} and in a wider context of generating samples and exploring the latent space of machine learning models \cite{fontaine_illuminating_2021, hagg_efficient_2023, chandra_bayesian_2024}.

\subsection{Extension to Other Systems}
To demonstrate that the method can generalise to other systems, three material systems were selected: C, \ce{SiC} and \ce{SiO2}. 
These are again known to adopt multiple polymorphs.
No hyperparameter adjustments were made to improve performance on the individual systems. 
The only changes were the adjustment of the feature vector limits to capture the values of the corresponding references, and the allowed random symmetries used by \texttt{pyxtal} to generate random structures.

The results are summarised in Table \ref{tab:other_materials_stats} below. 
For all materials tested above, between 55\% (\ce{C}) and 77\% (\ce{SiO2}) of structures available to be discovered were found. 
The absolute number of structures identified could likely be larger, but the resolution of the grid creates a cap. 
For instance, the 99 reference structure of \ce{SiO2}, are distributed across only 28 centroids (Figure available in ESI). 
Therefore, inherently due to the resolution of the grid, we are unlikely to find all reference structures. 
The ground state was found 7 out of 10 times for C and it was not found in any experiments for \ce{SiO2}. 
This decreased performance as compared to \ce{TiO2} is expected, as the parameters of the experiment were not adjusted. 
\ce{SiC} was excluded from this analysis as its ground state structure contains more than the 24 atoms available for optimisation.

\begingroup
    
\renewcommand{\arraystretch}{1.5} 
\begin{table}
\small
    \caption{Summary statistics on C, \ce{SiO2} and \ce{SiC} averaged on 10 experiments. NB: The ground state of Silicon Carbide has more than 24 atoms therefore it was not available to be found.}
    \label{tab:other_materials_stats}
    \begin{center}
    \begin{tabular*}{0.48\textwidth}{@{\centering\extracolsep{\fill}}
    p{0.4\linewidth} p{0.15\linewidth} p{0.15\linewidth} p{0.15\linewidth} 
    }
    \hline
         & Carbon & Silicon Dioxide & Silicon Carbide\\
    \hline
        Unique matches ($\sigma$) & 15 (2.4) & 20 (3.4) & 6 (0.48) \\ 
        Unique gold standard / high confidence matches ($\sigma$) & 6 (1.3) & 3 (1.4) & 3 (0)\\ 
        Reference structures (number of filled centroids) & 44 (26) & 99 (28) & 13 (11)\\
        Number of times reference ground state found & 7 & 0 & N/A\\
    \hline
    \end{tabular*}
    \end{center}

\end{table}

\endgroup

\section{Discussion \& Further Work}
By combining the strengths of MAP-Elites with the framework of evolutionary algorithms used in crystal structure prediction, we built a pipeline that can be used in a range of applications to identify diverse crystal structures.
This work assumes almost no prior knowledge of the material system making it highly generic.
Although we focused on polymorphs of single compositions, this method could be expanded to explore larger chemical spaces. 
Additionally, since the feature vector is defined by the user, this method can be applied out-of-the-box in a wide range of applications. 

The underlying evolutionary algorithm uses simple mutations and makes limited assumptions regarding the crystal system. 
As such, we expect that by implementing a more advanced procedure under-the-hood of MAP-Elites, such as improved starting structures and tailored mutation operators, this technique could yield promising results on larger and more complex systems including artificial heterostructures.

One of the requirements of MAP-Elites is a high number of required evaluations. 
As evaluating thousands of structures using DFT is not feasible, surrogate models are beneficial. 
These, however, rely on the data available for training. 
To fine-tune models in particular search spaces, MAP-Elites could be used to identify structures outside of the training distribution that fool the model.
We observed the ability of the algorithm to identify structures that exploit the underlying interatomic model to find structures with higher absolute energy values than expected, thus making it suitable for such an application.

Building on this work, extensions of QD techniques could also be used. For instance, multi-objective QD could be employed to enable the search for materials with conflicting objectives, while while discovering large collections of structures that span across the property space\cite{pierrot_multi-objective_2022}.
Alternatively, if differentiable models are used for all feature and fitness function models, gradients can be used to inform the mutations thus allowing solutions to converge faster. This is done in Differentiable Quality-Diversity \cite{fontaine_differentiable_nodate}.

\section{Conclusion}
We presented the application of Quality-Diversity algorithms to the problem of crystal structure prediction. 
By using properties of materials to discretise the search space and the \ce{TiO2} system we demonstrated that this technique can not only find the ground state structure but other structures with varying properties. 
The versatility of this algorithm was then validated on three other crystal systems, where novel structures were found for C, \ce{SiO2} and \ce{SiC}. 
The performance of our method can benefit from improvements to the underlying algorithm to combine state-of-the-art techniques from evolutionary algorithms applied to crystal structure prediction with QD algorithms to reap the benefits of faster convergence to realistic structures with a diversity of results. 

\section*{Data availability statement}
A repository containing the data and associated analysis code have been made available on Github (\url{https://github.com/adaptive-intelligent-robotics/QD4CSP}).

\section*{Author Contributions}
M.W., A.C and A.W. designed the study. 
M.W. and A.C. analysed the results and discussed additional experiments from a computational perspective.
M.W and A.W. analysed results from a materials science perspective. 
M.W. conducted the experiments, established the methodology, wrote the software and wrote the original draft of the paper. 
A.C supervised the work.
A.C and A.W reviewed and edited the paper.

\section*{Conflicts of interest}
There are no conflicts to declare.

\balance

\renewcommand\refname{References}

\bibliography{references_manual} 
\bibliographystyle{rsc} 

\end{document}



\begin{center}
    {\noindent \centering \LARGE \textbf{Electronic Supplementary Information: \newline Illuminating the property space in crystal structure prediction using Quality-Diversity algorithms}}

\end{center}

\section{Hyperparameter Selection}

The hyperparameters used in this work are based on previous work in the field as summarised below:

\noindent
\begin{table}[H]
\footnotesize
\centering
\begin{tabular}{|p{2.5cm}|p{2.5cm}| p{7cm}| p{1.5cm}|}
\hline
\textbf{Parameter} & \textbf{Value}  & \textbf{Reason} & \textbf{Reference} \\ \hline
Number of atoms & 24 & The number used in the references is 48, because increasing the number of atoms increases the difficulty to find the ground state. However to balance the trade-off between difficulty and time required for an experiment a smaller size of 24 atoms was selected for this investigation. & \cite{falls_xtalopt_2021, lyakhov_new_2013, lonie_xtalopt_2011} \\ \hline
Unit cell parameter limits & unit cell vectors: $2-60$ \AA;  angles:  $0-\pi$  & These were set to be less restrictive than the already considered 'loose' parameters $0- 40$\AA, and $20-120$, degrees used by Lonie and Zurek. & \cite{lonie_xtalopt_2011} \\ \hline
Maximum angle between unit cell vector and plane created by other 2 vectors & [20, 160] , [60, 120], [ 20, 160] & As above, this was set to be less restrictive than the example provided in the reference implementation of a genetic algorithm. & \cite{van_den_bossche_tight-binding_2018, hjorth_larsen_atomic_2017}  \\ \hline
Unit cell volume & 450 \AA$^3$. &  This was estimated based on the  $483$ \AA$^3$ used by Lonie and Zurek in the reference. & \cite{lonie_xtalopt_2011} \\ \hline
Scaling factor between atomic radii & 0.4 &  This parameter defines how closely atoms can be positioned in the unit cell. Set to the values used in references. & \cite{falls_xtalopt_2021, lyakhov_new_2013}. \\ \hline
Crossover operators and their probabilities & Operators: Strain and Permutation mutations with probabilities: 50\%, 50\%  &  The initial set up included crossover and soft mutation operators as set by Lykahov et al in the reference. These were not included because they did not guarantee an individual to be outputted and soft mutation also requires realistic individuals which could not be guaranteed in this work. Therefore for computational efficiency these were not included.  & \cite{lyakhov_new_2013} \\ \hline
Unit cell splitting factors & 2, 4 &  This is a parameter used to determine transnational symmetry when using the random structure generator. The values were set using the reference & \cite{lyakhov_new_2013} \\ \hline
Starting Population Size & 20 &  The size used by the Lyakhov et al. is 10, but since our objective is also around diversity a larger starting point was used. This and larger population sizes were also tested as reported in table \ref{tab:base_hyperparams}. as part of the MAP-Elites parameters search & \cite{lyakhov_new_2013} \\ \hline

\end{tabular}
\caption{Hyperparameter selection for \ce{TiO2} mapped to past work. }
\end{table}

Additionally a brief summary of the hyperparameters set from a computational perspective are summarised below:

\begin{table}[H]
\footnotesize
\centering
\begin{tabular}{|p{3cm}|p{1.5cm}|p{1.5cm}| p{8cm}|}
\hline
\textbf{Parameter} & \textbf{Value Set} & \textbf{Values Tested} & \textbf{Reason} \\ \hline
Number of niches & 200                         & 200, 500         & 200 niches was selected to increase the chances of competition between structures. The dynamics of the experiment were consistent across the three experiments so 200 was selected as a good trade off between speed and diversity of possible solutions. \\ \hline
Proportion of niches to be filled                                                       & 0.1                         & 0.1, 0.2               & The higher proportion of niches didn't improve the average fitness obtained nor the number of reference matches found, so the lower threshold was selected to move to structure mutations quicker.                                                      \\ \hline
Number of random structures to initialise                                               & 20                          & 20, 40, 80            & 20 was set as it limited the time spent on generating new structures, and if required would be called multiple times to filled the proportion of niches desired so on its own this parameter had limited impact.   \\ \hline
Batch size per generation  & 100                         & 20, 50, 100                & This parameter defines how many individuals will be mutated per generation. Setting this to 100 meant that some individuals were mutated multiple times. This accelerated finding of reference matches.                                                   \\ \hline
\end{tabular}
\caption{Default hyperparameter selection.}
\label{tab:base_hyperparams}
\end{table}

\newpage

\section{Selection of force threshold value}
\begin{figure}[h]
    \centering
    \includegraphics[width=0.5\textwidth]{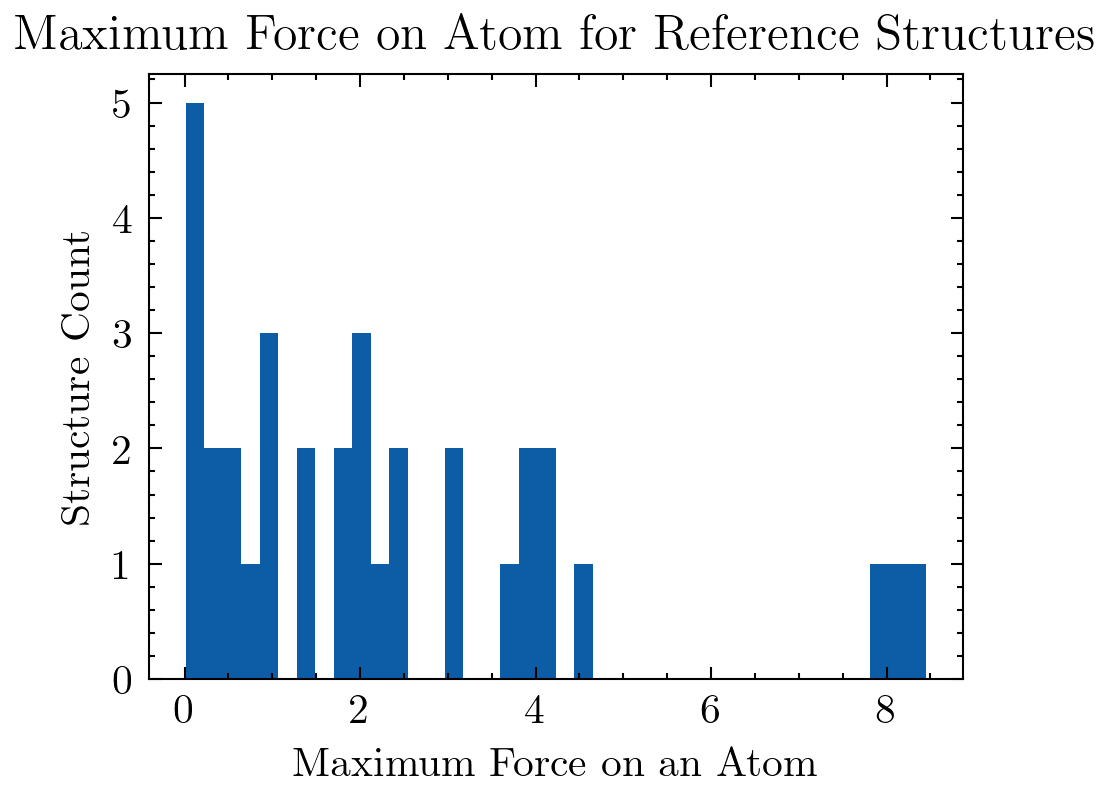}
    \caption{Histogram of the maximum force acting on each of the reference structures of \ce{TiO2} computed using \texttt{CHGNet}\cite{deng_chgnet_2023}.}
    \label{fig:fmax}
\end{figure}

The maximum force acting on each of the reference structures was computed. 
This includes the forces acting on each atom as well as the stresses on the unit cell, as implemented and used for relaxation in \texttt{CHGNEt}. 
The results are presented in the histogram in Figure \ref{fig:fmax}.

\section{Identifying equivalent structures}
Within the tolerances set in this work we observed that some structures would be equivalent to each other. 
This is demonstrated within Figure \ref{fig:ref_struct_match} below. 
There the confusion matrix was constructed by comparing all reference structures to each other. 
Structures considered equivalent to each other are marked with green.
In Figure \ref{fig:ref_struct_match_all} we can observe that equivalent pairs are: 
\textit{mp-390} and \textit{mp-34688},
\textit{mp-2657} and \textit{1041565}.

\begin{figure}[H]
    \centering
    \begin{subfigure}[b]{0.35\textwidth}
    \centering
        \includegraphics[width=\textwidth]{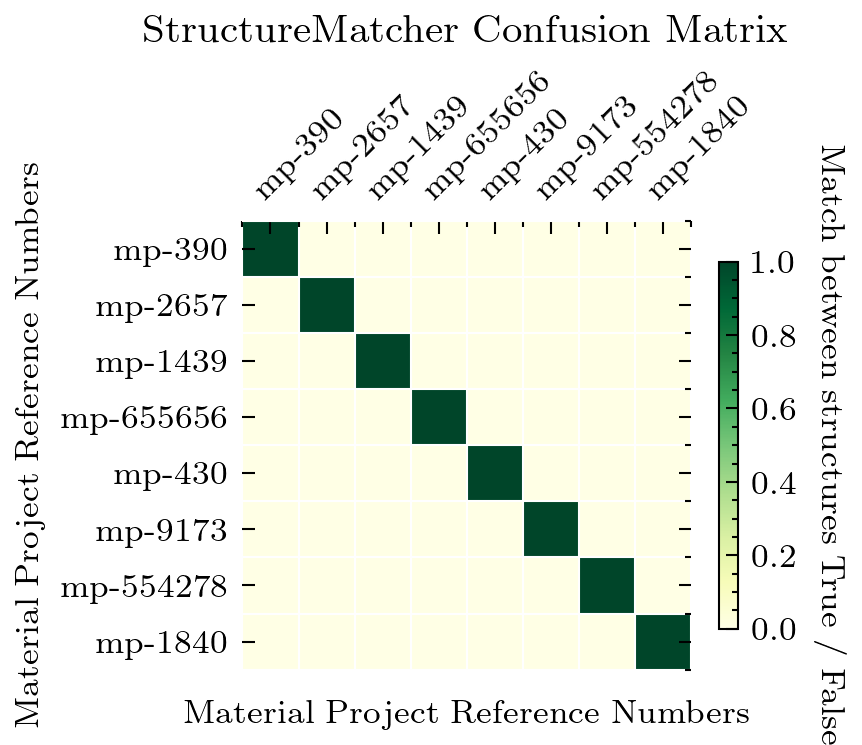}
        \caption{Matches between experimentally observed reference structures.}
        \label{fig:ref_struct_match_experimental}
    \end{subfigure}
    \hspace{1cm}
    \begin{subfigure}[b]{0.55\textwidth}

    \centering
        \includegraphics{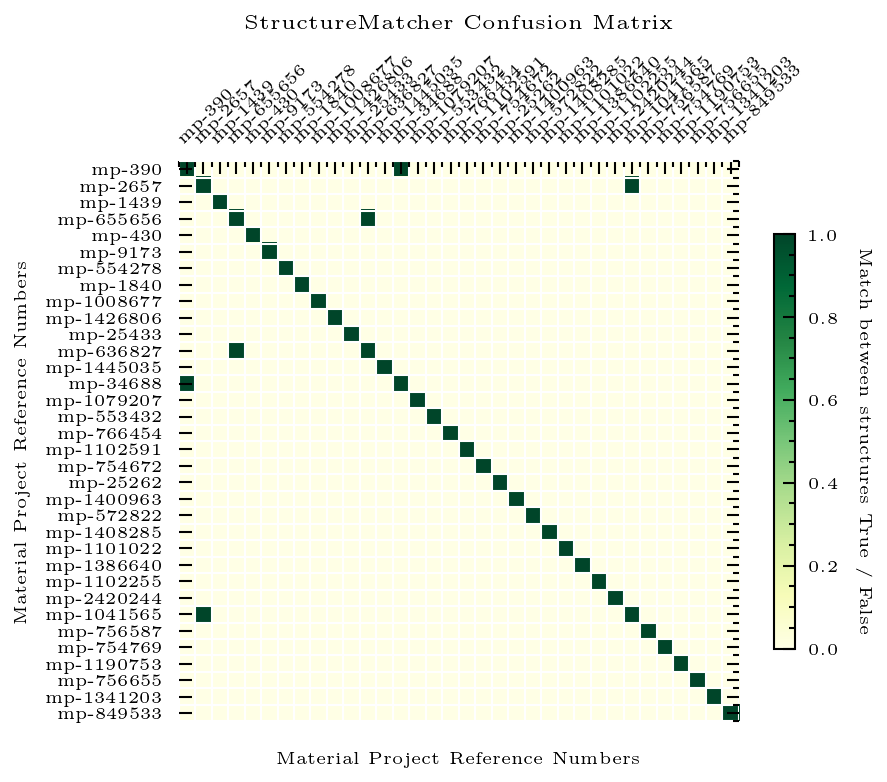}
        \caption{Matches between all reference structures.}
        \label{fig:ref_struct_match_all}
    \end{subfigure}%
    \caption{Correlation plot of matches between reference structures as generated using \texttt{StructureMatcher} from \texttt{pymatgen}\cite{ong_python_2013}.}
    \label{fig:ref_struct_match}
\end{figure}

\section{Sample results for \ce{C}, \ce{SiC}, \ce{SiO2}}
Below in Figure \ref{fig:validation_materials} we demonstrate sample archives for \ce{C}, \ce{SiC} and \ce{SiO2}. 
We can observe similar trend as with \ce{TiO2}; the archives largely developed in the expected areas of the features space. 
With the exception of \ce{SiC} the archives demonstrate that a wide set of solutions is demonstrated. 

\begin{figure}
    \centering
    \includegraphics[height=\textheight]{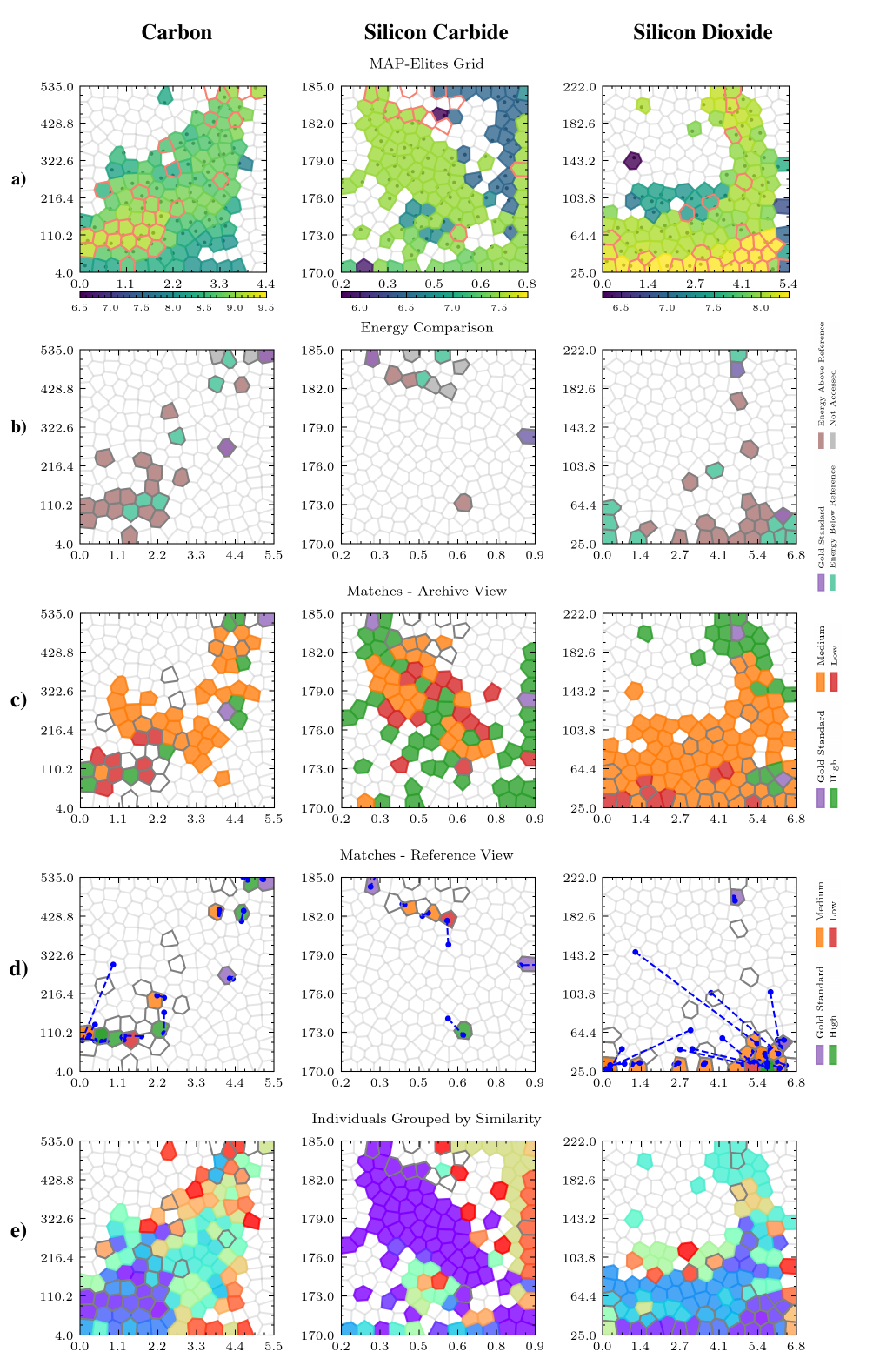}
    \caption{Sample results for \ce{C}, \ce{SiC} and \ce{SiO2} experiments.}
    \label{fig:validation_materials}
\end{figure}

\newpage
\bibliographystyle{rsc} 
\bibliography{references_manual} 